\documentclass[conference]{IEEEtran}
\IEEEoverridecommandlockouts
\usepackage{cite}
\usepackage[table,xcdraw]{xcolor}
\usepackage{amsmath,amssymb,amsfonts}
\usepackage{algorithmic}
\usepackage{textcomp}
\usepackage{framed}
\usepackage [utf8]{inputenc}
\usepackage [T1] {fontenc}
\usepackage {hyperref}
\usepackage {url}
\usepackage{tikz}

\usepackage {amsfonts}

\usepackage {nicefrac}

\usepackage {microtype}

\usepackage {enumitem}

\usepackage{graphicx}
\usepackage{amsmath}
\usepackage{amssymb}
\usepackage{multirow}
\usepackage{array}
\usepackage{makecell}

\usepackage{booktabs}

\begin{document}

\title{A Cascaded Dilated Convolution Approach for  Mpox Lesion Classification\\
}

\author{\IEEEauthorblockN{Ayush Deshmukh\thanks{Ayush Deshmukh is with the department of Information Technology at the National Institute of Technology, Karnataka, India (e-mail: aayush.m.deshmukh@gmail.com, aad.211it014@nitk.edu.in).}
\thanks{The code will be available \href{https://github.com/TheUndercover01/CAGA-A-Cascaded-Dilated-Convolution-Approach-for-Mpox-Lesion-Classification/tree/main}{here.}}}}

\maketitle

\begin{abstract}

The global outbreak of the Mpox virus, classified as a Public Health Emergency of International Concern (PHEIC) by the World Health Organization, presents significant diagnostic challenges due to its visual similarity to other skin lesion diseases. Traditional diagnostic methods for Mpox, which rely on clinical symptoms and laboratory tests, are slow and labor intensive. Deep learning-based approaches for skin lesion classification offer a promising alternative. However, developing a model that balances efficiency with accuracy is crucial to ensure reliable and timely diagnosis without compromising performance. This study introduces the Cascaded Atrous Group Attention (CAGA) module to address these challenges, combining the Cascaded Atrous Attention and the Cascaded Group Attention mechanisms. The Cascaded Atrous Attention module utilizes dilated convolutions and cascades the outputs to enhance multi-scale representation. This is integrated into the Cascaded Group Attention mechanism, which reduces redundancy in Multi-Head Self-Attention. By integrating the Cascaded Atrous Group Attention module with EfficientViT-L1 as the backbone architecture, this approach achieves state-of-the-art performance, reaching an accuracy of 98\% on the Mpox Close Skin Image (MCSI) dataset while reducing model parameters by 37.5\% compared to the original EfficientViT-L1. The model's robustness is demonstrated through extensive validation on two additional benchmark datasets, where it consistently outperforms existing approaches.

\end{abstract}


\section{Introduction}

The recent global outbreak of the Mpox virus has posed significant challenges for public health authorities and researchers worldwide \cite{WHO2022}\cite{ :/content/10.2807/1560-7917.ES.2023.28.12.2200772}. As this zoonotic disease rapidly spread beyond endemic regions, it affected multiple non-endemic countries, including the Netherlands \cite{:/content/10.2807/1560-7917.ES.2023.28.12.2200772}, Greece \cite{v15061384}, Portugal \cite{borges2023viral}, and France \cite{:/content/10.2807/1560-7917.ES.2023.28.50.2200923}. To date, the outbreak has infected over 100,000 individuals across hundreds of countries \cite{owid-mpox}. This surge has made the need for effective detection and diagnostic methods increasingly urgent. In response to this escalating crisis, the World Health Organization (WHO) declared Mpox a Public Health Emergency of International Concern (PHEIC).

Traditional approaches relying on clinical symptoms and laboratory testing have proven slow and laborious in the face of this evolving outbreak \cite{silva2023clinical}\cite{vaccines11061093}. This challenge is compounded by the clinical presentation of Mpox, which shares notable similarities with other skin lesion-based diseases, such as Chickenpox and Measles. Like these conditions, Mpox can manifest with a range of skin lesions, including rashes and vesicular eruptions, making accurate differentiation difficult. The resemblance of Mpox lesions to those caused by other infectious diseases, coupled with the limitations of existing diagnostic techniques, such as reporting delays and challenges in identifying transmission sources, has significantly impacted the precision and timeliness of diagnosis. These challenges underscore the urgent need for improved, timely, and accurate detection methods to contain the outbreak and mitigate its public health impact \cite{prasad2023dermatologic}.

In addressing these diagnostic challenges, maintaining high accuracy while improving efficiency is paramount. Any new diagnostic approach must achieve both speed and precision, as misdiagnosis could lead to inappropriate treatment protocols and ineffective containment measures. The stakes are exceptionally high, given the public health implications of incorrect or delayed diagnoses.

In this context, the rapid advancements in artificial intelligence, particularly in Convolutional Neural Networks (CNNs), offer a promising solution. Over the past decade, CNNs have demonstrated remarkable success in skin lesion classification\cite{esteva2017dermatologist}\cite{haenssle2018man}\cite{thomsen2020systematic}\cite{7893267}, paving the way for their potential application in improving Mpox detection and management\cite{chadaga2023application}\cite{asif2024ai}.

Although progress has been made in the identification of infectious diseases through medical imaging techniques, research on Mpox remains limited. These investigations have primarily concentrated on three critical methodological domains: adapting pre-trained medical imaging models\cite{jaradat2023automated}\cite{CAMPANA2024101874}, developing specialized neural network structures\cite{10491259}, and implementing advanced feature extraction techniques to capture the unique morphological characteristics of Mpox skin lesions\cite{thieme2023deep}.

While early applications of deep learning to Mpox detection show promise, they face several critical methodological challenges. One major issue is the considerable variation in model validation, with many studies relying on evaluations using only a single dataset. A model that performs well on one dataset may not deliver similar results on others, which limits its generalizability and reduces its reliability when applied to diverse real-world data. 

Moreover, comparative studies often use flawed assessment strategies, such as comparing attention-based models with a narrow set of non-attention-based architectures. These limited comparisons undermine the scientific reliability of performance assessments in Mpox detection, as they fail to account for a wide range of potential models. Without considering a broader set of architectures, the results can present a skewed view of a model's true effectiveness, leading to inaccurate conclusions about which approaches are most suitable for Mpox detection in real-world scenarios. These challenges, including the lack of large-scale, standardized benchmark datasets, have slowed the development of reliable models for Mpox detection, widening the gap between research and clinical application and raising concerns about the trustworthiness of current computational diagnostic methods.

%The development of such models must also carefully balance computational efficiency with diagnostic accuracy. While faster processing times are desirable, they cannot come at the cost of reduced accuracy, as reliable diagnosis remains the primary goal in clinical settings. 

A critical gap exists in Mpox classification models that are both rigorously tested and strike a balance between computational efficiency and diagnostic accuracy. While reducing processing times is important, it should never compromise the accuracy of the diagnosis, as reliable diagnosis remains the primary goal in clinical settings.
%Despite the demonstrated effectiveness of transformer-based architectures in skin lesion classification \cite{zhang2019attention}\cite{8710336}, their application to Mpox detection has been minimal. Most existing studies focus predominantly on traditional Convolutional Neural Networks , leaving a significant gap in leveraging the capabilities of transformer models for this specific task.

\begin{figure}
\includegraphics[width=\columnwidth]{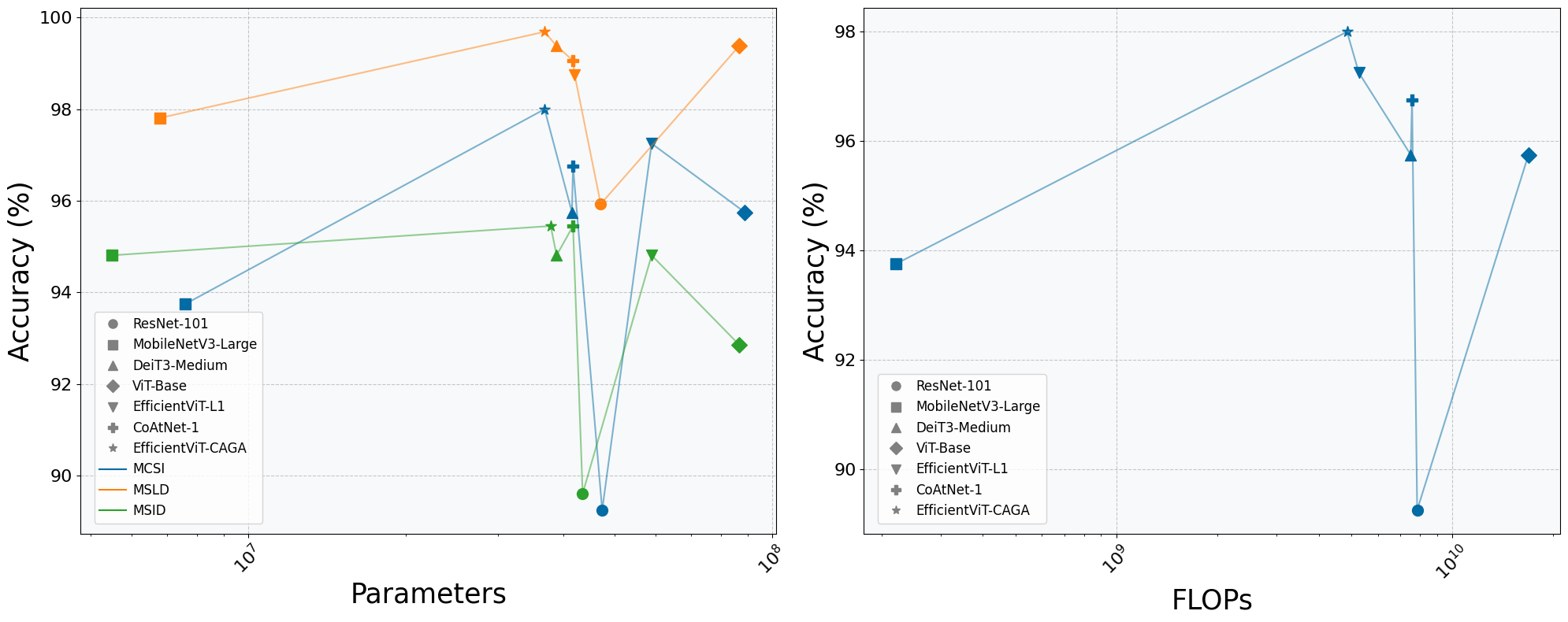}
\caption{Parameter and accuracy comparison (Left) and FLOPs and accuracy comparison (Right). \textit{Left:} Comparison between EfficientViT-CAGA (proposed) and other CNN and ViT models across three datasets: MCSI, MSID, and MSLD, with mean accuracy shown across 10 folds. \textit{Right:} Comparison on the MCSI dataset, showing the mean accuracy across 10 folds.}
\label{fig:efficiency_metrics}
\vspace{-10pt}
\end{figure}

The main contributions of this paper are thus summarized
as follows:
\begin{enumerate}
    \item I propose a novel Cascaded Atrous Group Attention (CAGA) module, which consists of a unique Cascaded Atrous Attention (CAA) integrated within a Cascaded Group Attention (CGA) structure. This module leverages dilations to capture multi-scale information, self-attention to capture long-range dependencies, and a cascading structure to enhance feature flow across dilations and attention heads.
    \item I demonstrate that by integrating the novel CAGA module with the EfficientViT-L1 backbone \cite{cai2024efficientvitmultiscalelinearattention}, the proposed model achieves state-of-the-art (SOTA) accuracy across all evaluated datasets while also maintaining impressive computational efficiency. To assess the effectiveness of this model, I conducted extensive evaluations on three diverse datasets: Mpox Close Skin Image Dataset (MCSI)\cite{CAMPANA2024101874}, Monkeypox Skin Images Dataset (MSID)\cite{bala2023monkeynet}, and Monkeypox Skin Lesion Dataset (MSLD)\cite{Nafisa2022}\cite{Nafisa2023}. As shown in Fig. \ref{fig:efficiency_metrics}, the proposed EfficientViT-CAGA model not only achieves the highest accuracy across all datasets but also does so while requiring fewer parameters and FLOPs compared to most other state-of-the-art models.
    
    \item Through comprehensive ablation studies, I validate the effectiveness of CAGA's design choices and further provide visual interpretability through Grad-CAM analysis to demonstrate how the proposed model attends to relevant features. 
\end{enumerate}

\section{Related Works}

\subsection{Vision Transformers}

% \begin{figure}
%     \centering
%     \begin{minipage}{\linewidth}
%         \begin{minipage}[b]{0.485\linewidth}
%             \centering
%             \includegraphics[width=\linewidth]{params.png}
%             %\caption{Accuracy vs FLOPs}
%             \label{fig:flops}
%         \end{minipage}%
%         \hfill%
%         \begin{minipage}[b]{0.485\linewidth}
%             \centering
%             \includegraphics[width=\linewidth]{flops.png}
%             %\caption{Accuracy vs Parameters}
%             \label{fig:params}
%         \end{minipage}
%     \end{minipage}
%     %\vspace{-10pt}  % Adjust this value as needed
%     \caption{Parameter and accuracy comparison (Left) and FLOPs and accuracy comparison (Right). \textit{Left:} Comparison between EfficientViT-CAGA (proposed) and other CNN and ViT models across three datasets: MCSI, MSID, and MSLD, with mean accuracy shown across 10 folds. \textit{Right:} Comparison on the MCSI dataset, showing the mean accuracy across 10 folds.}
%     \vspace{-10pt}  % Adjust this value as needed
%     \label{fig:efficiency_metrics}
% \end{figure}
Convolutional Neural Networks (CNNs) have been instrumental in advancing the field of computer vision \cite{szegedy2014goingdeeperconvolutions}\cite{ NIPS2012_c399862d}\cite{xie2017aggregatedresidualtransformationsdeep}\cite{he2015deepresiduallearningimage}\cite{chollet2017xceptiondeeplearningdepthwise}\cite{tan2020efficientnetrethinkingmodelscaling}. In recent years, Transformers \cite{vaswani2023attentionneed} have achieved remarkable success in natural language processing (NLP) tasks. Inspired by this success, researchers have explored the application of Transformers to Computer Vision, with one of the most influential efforts being the Vision Transformer (ViT)\cite{dosovitskiy2021imageworth16x16words}.

ViT was one of the first pure Transformer-based models to achieve state-of-the-art results on ImageNet. Unlike CNNs, which rely on local receptive fields and other inductive biases, ViT processes images as sequences of patches, enabling a more global understanding of the visual content. However, their lack of inherent inductive biases, such as locality, makes them computationally intensive and heavily dependent on large-scale data\cite{dosovitskiy2021imageworth16x16words}\cite{raghu2022visiontransformerslikeconvolutional}\cite{park2022visiontransformerswork}\cite{chen2022visiontransformersoutperformresnets}. Hybrid architectures\cite{guo2022cmtconvolutionalneuralnetworks}\cite{liu2022convnet2020s}\cite{graham2021levitvisiontransformerconvnets}\cite{liu2021swintransformerhierarchicalvision}\cite{dai2021coatnetmarryingconvolutionattention} seek to balance these trade-offs, integrating convolutional operations to model local features while leveraging attention mechanisms for global context. This study employs a hybrid architecture, combining the convolutional backbone of EfficientViT-L1 with the Cascaded Atrous Group Attention mechanism for self-attention, resulting in significantly improved computational efficiency and performance.

\subsection{Deep Learning for Mpox Classification}

Deep learning has shown promise for Mpox classification, offering potential improvements in early diagnosis and outbreak control. As noted in \cite{WONGVIBULSIN2023283}, these algorithms are effective in diagnosing Mpox from skin lesions but still face challenges in development and implementation.

 Thieme et al.\cite{thieme2023deep} created a dataset of 139,198 skin lesion images, with Mpox and non-Mpox cases. They developed the MPXV-CNN, a deep convolutional neural network that demonstrated high sensitivity and specificity in validation and testing, and introduced PoxApp, a web-based tool that facilitates its use in patient management. Raha et al. \cite{10491259} presented a novel approach for Mpox detection using an attention-powered MobileNetV2 architecture. The authors emphasized the importance of interpretability in deep learning models for medical image analysis. They compared their proposed model with several baseline models, including VGG-19, ResNet-152, MobileNetV2, AlexNet, GoogleLeNet, and ShuffleNetV2. The study employed a comprehensive set of evaluation metrics and utilized LIME and Grad-CAM techniques for model interpretability. They also outlined a detailed operational workflow for early Mpox detection, covering steps from image capture to decision support. Jaradat et al.\cite{jaradat2023automated} evaluated the performance of various pre-trained models, including VGG16, VGG19, ResNet50, MobileNetV2, and EfficientNetB3, on a dataset of 117 skin lesion images. The authors found that data augmentation techniques helped improve the model’s performance significantly. Asif et al.\cite{ASIF2023342} proposed an ensemble learning approach for Mpox detection, combining three pre-trained CNN models (DenseNet201, MobileNet, and DenseNet169) and optimizing the weights using the Particle Swarm Optimization (PSO) algorithm. This ensemble-based approach achieved higher classification accuracy compared to the individual models.

 \begin{figure*}[t!]
\includegraphics[width=180mm]{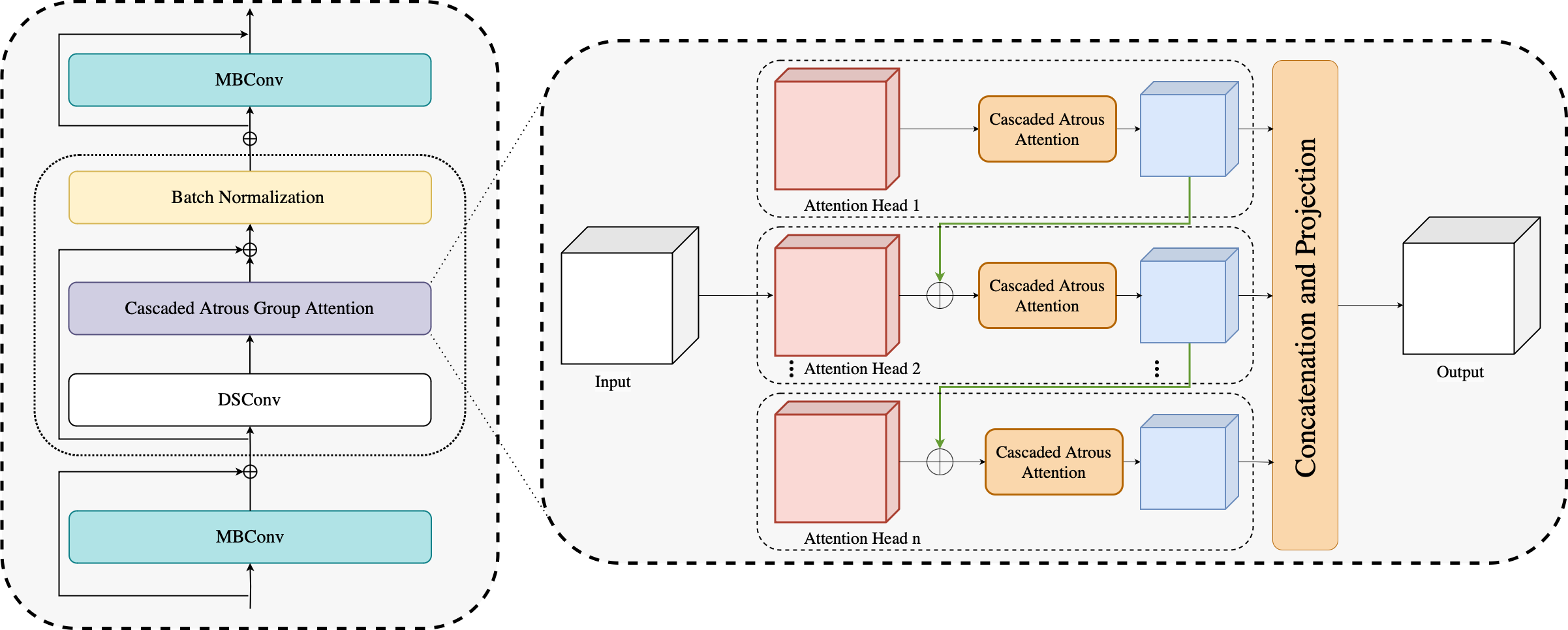}
\caption{Integration of CAGA within the Network Architecture (left) and detailed structure of Cascaded Group Attention with Cascaded Atrous Attention Module (right). \textit{Left:} The MBConv layer represents the Mobile Inverted Bottleneck Convolution\cite{sandler2019mobilenetv2invertedresidualslinear}, which is part of the EfficientViT-L1 architecture. A Depthwise Separable Convolution\cite{chollet2017xceptiondeeplearningdepthwise} is used to reduce computational complexity while preserving spatial and channel-wise feature interactions, converting the input to an $n \times h$ dimension. Batch Normalization (BN)\cite{ioffe2015batchnormalizationacceleratingdeep} is applied to the output of the residual connection.}
\label{fig: CAGA}

\end{figure*}

\begin{figure*}[t!]
\includegraphics[width=180mm]{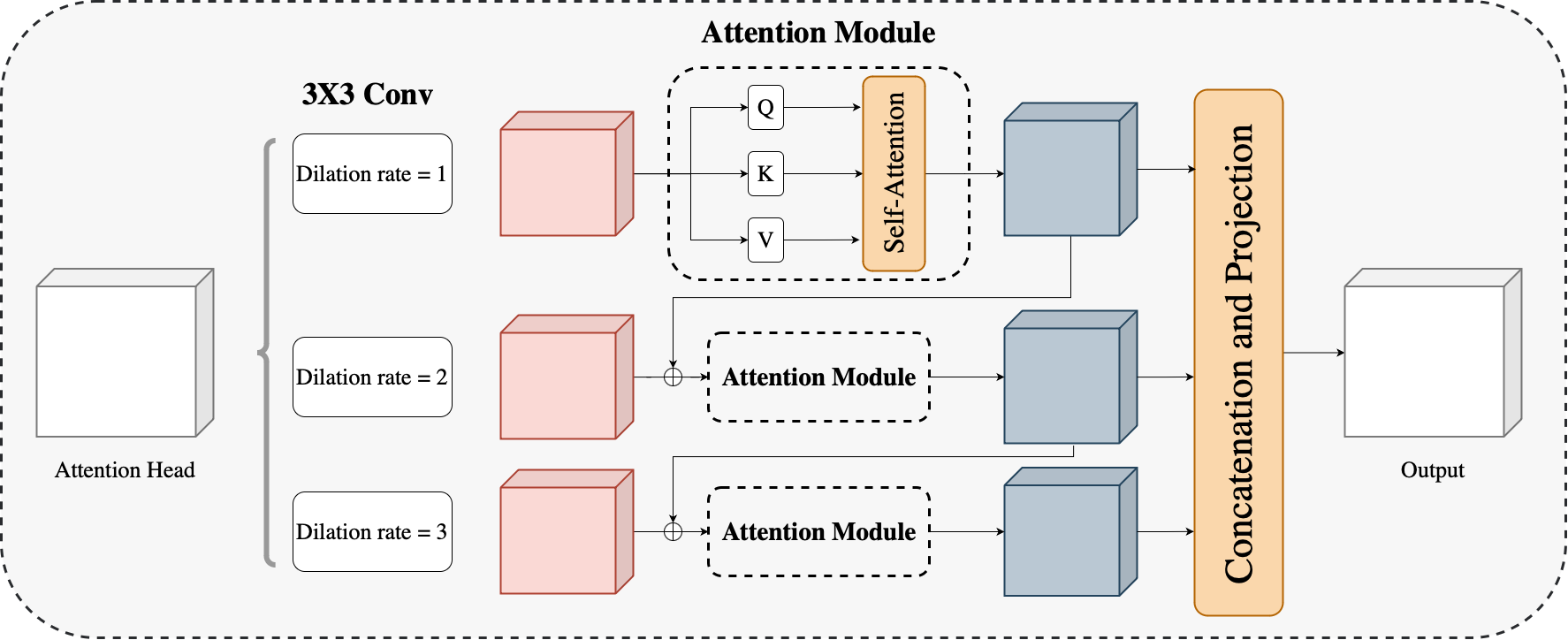}
\caption{Cascaded Atrous Attention. }
\label{CAA}
\vspace{-10pt}
\end{figure*}

These works in Mpox detection highlight a trade-off between model complexity and performance, often relying on resource-intensive ensembles or simplified architectures. This study addresses the core challenge of efficient feature extraction with the novel Cascaded Atrous Group Attention mechanism, achieving high accuracy across multiple datasets while remaining computationally efficient.

\section{Methodology}
This section presents the Cascaded Atrous Group Attention. I first provide a comprehensive overview of the module, followed by detailed explanations of its key components: the Cascaded Atrous Attention (CAA) and the Cascaded Group Attention (CGA) mechanisms. 

\subsection{Cascaded Atrous Group Attention (CAGA)}

I introduce Cascaded Atrous Group Attention (see Fig. \ref{fig: CAGA} (right)) an architecture that builds upon the Cascaded Group Attention (CGA)\cite{liu2023efficientvitmemoryefficientvision} module. While the CGA directly processes each attention head using standard self-attention, this work replaces this mechanism with the proposed Cascaded Atrous Attention (CAA). By integrating multi-scale dilated convolutions and a sophisticated self-attention mechanism for each dilation, CAA enhances the feature representation capabilities of the CGA approach.

\subsubsection{Cascaded Atrous Attention (CAA)}

\begin{table*} [ht]
\centering

\caption{Comparison with SOTA Image Classification Models on MSID Dataset}

\setlength{\tabcolsep}{7pt}
\renewcommand{\arraystretch}{1.3}
\begin{tabular}{l|ccccccccc}
\toprule
\multicolumn{1}{l}{Architectures} &
\multicolumn{1}{c}{Parameters} &

\multicolumn{1}{c}{Accuracy} &
\multicolumn{1}{c}{Precision} &
\multicolumn{1}{c}{Recall} &
\multicolumn{1}{c}{F1 Score} &
\multicolumn{4}{c}{Classwise Accuracy} \\
\cmidrule(lr){7-10}
\multicolumn{6}{c}{} & Chickenpox & Measles & Mpox & Normal \\
\midrule
ResNet-101           & 43.5M &  0.8961     & 0.8633        & 0.8782    & 0.8686 & 0.762     & 0.8333        & 0.875             & \underline{0.9831}\\
MobileNetV3-Large     & \textbf{5.5M} & \underline{0.9481} & \underline{0.94}  & 0.927 & \underline{0.9327} & \underline{0.9047}     & \textbf{0.9444}        & 0.9107             & \textbf{1.0}\\
DeiT3-Medium         & 38.8M &  \underline{0.9481} & 0.9159  & 0.9308 & 0.9229 & 0.8095    & \underline{0.8889}       & \textbf{0.9821}              & \underline{0.9831}\\
ViT-Base                  & 86.5M & 0.9286 & 0.9097  & 0.9154 & 0.9113  & 0.8571     & \underline{0.8889}       & 0.893             & \textbf{1.0}\\
EfficientViT-L1          & 58.9M &  \underline{0.9481} & 0.9082 & 0.9273 & 0.9165 & 0.761     & \underline{0.8889}        & \textbf{0.9821}             & \textbf{1.0}\\
CoAtNet-1              & 41.7M &  \textbf{0.9545} & 0.9201  & \textbf{0.9454} & 0.9317 & 0.8095     & \underline{0.8889}       & \textbf{0.9821}             & \textbf{1.0} \\
\textbf{EfficientViT-CAGA }           & \underline{37.8M} & \textbf{0.9545} & \textbf{0.9523} & \underline{0.9376}  & \textbf{0.9446} & \textbf{0.9524}    & \textbf{0.9444}      & \underline{0.9464}     & 0.9661\\
\bottomrule
\end{tabular}
\label{table:msid}

\end{table*}

Inspired by DeepLab's Atrous Spatial Pyramid Pooling (ASPP) \cite{deeplab}, I propose a multi-scale feature extraction module called Cascaded Atrous Attention (CAA). This module leverages parallel dilated convolutions to effectively capture contextual information at multiple receptive fields. Building upon ASPP's use of varying dilation rates, CAA introduces a cascaded approach that combines dilated inputs with a self-attention mechanism for enhanced feature representation. Fig. \ref{CAA} provides an illustration of the proposed CAA module. 

I incorporate dilation rates \( d \in \{1,2,3\} \) applied to convolution operations with kernel size \( k \times k \) (typically \( 3 \times 3 \)) across each attention head \( X_{i} \in \mathbb{R}^{h \times H \times W} \), where \( (H, W) \) is the resolution, \( h \) denotes the dimensionality of the attention head, and \( 1 \leq i \leq n \), where \( n \) is the total number of heads. Which is followed with a $1\times1$ convolution to transform the feature map into appropriate dimensionality required for self attention. Let \( \tilde{H} = \frac{H - k_{\text{eff}} + s}{s} \) and \( \tilde{W} = \frac{W - k_{\text{eff}} + s}{s} \), then the output after dilated convolution is denoted as
\( X_h^{d} \in \mathbb{R}^{(3 \times d_{qkv}) \times \tilde{H} \times \tilde{W}} \), 
where \( k_{\text{eff}} = k + (k-1)(d-1) \) represents the effective kernel size. 
Here, \( d_{qkv} \) is the dimension of query, key, and value embeddings, and \( s \) is the stride. 
%, where \((H, W)\) is the resolution and \(h\) denotes the dimensionality of the attention head.
The resulting feature map is split into \( Q^d, K^d, V^d \in \mathbb{R}^{d_{qkv} \times (\tilde{H} \times \tilde{W})} \) and passed to the self-attention module. The self-attention then is computed as:

\begin{equation}
    \textit{Attn}_d = \text{Softmax}\left( \frac{Q^d (K^d)^T}{\sqrt{d_{qkv}}} \right) V^d
\end{equation}

 I introduce a hierarchical cascading approach to dilated self-attention maps that progressively aggregates and enhances spatial context. The model incrementally integrates proximal and broader neighborhood features, creating a multi-scale representation that captures nuanced spatial dependencies across different contextual granularities. To address dimensionality disparities between the dilated attention maps and subsequent dilated convolutional inputs, I implement a $1 \times 1$ convolution projection layer, thereby ensuring dimensional consistency.
\begin{equation}
    \tilde{\textit{X}}_{i}^{d} = \textit{X}_{i}^{d} + \text{Proj}(\textit{Attn}_{d-1}), \quad \ 1 < d \leq 3 
\end{equation}

Each dilated attention map is interpolated to restore the feature representation to its original spatial resolution of ($H, W$). These maps are then concatenated and subsequently projected back to match the attention head dimensionality.

\subsubsection{Cascaded Group Attention (CGA)}

Multi-Head Self-Attention (MHSA) suffers from attention head redundancy\cite{gao2019representationdegenerationproblemtraining} \cite{touvron2021goingdeeperimagetransformers}\cite{zhou2021deepvitdeepervisiontransformer}\cite{zhou2021refinerrefiningselfattentionvision}\cite{chen2022principlediversitytrainingstronger}, resulting in computational inefficiency. To address this limitation \cite{liu2023efficientvitmemoryefficientvision} propose Cascaded Group Attention, where the input \(X \in \mathbb{R}^{(n \times h) \times H \times W}\) is split into $n$ distinct heads of dimensionality $h$. Each attention head is processed independently, with a distinctive feature: for heads beyond the first, the input is augmented by adding the output of the preceding processed head \((\hat{X}_{i-1})\).

\begin{equation}
\tilde{\textit{X}}_{i} = \textit{X}_{i} + \hat{\textit{X}}_{i-1}, \quad \ 1 < i \leq n
\end{equation}

\begin{table*}[ht]
\centering
    
%\captionsetup{justification=centering}
\caption{Comparison with SOTA Image Classification Models on MCSI Dataset \\ Results Reported as Mean $\pm$ Standard Deviation (10-Fold Cross-Validation)}
\setlength{\tabcolsep}{7pt}
\renewcommand{\arraystretch}{1.3}
\begin{tabular}{l|cccccc}
\toprule
\multicolumn{1}{l}{Architectures} &
\multicolumn{1}{|c}{Parameters} &
\multicolumn{1}{c}{Accuracy} &
\multicolumn{1}{c}{Precision} &
\multicolumn{1}{c}{Recall} &
\multicolumn{1}{c}{F1 Score} \\
\midrule
ResNet-101          & 47.4M &  0.8925 $\pm$ 0.049 & 0.8908 $\pm$ 0.056 & 0.891 $\pm$ 0.059 & 0.884 $\pm$ 0.060 \\
MobileNetV3-Large   & \textbf{7.6M} &  0.9375 $\pm$ 0.054 & 0.9433 $\pm$ 0.052 & 0.9473 $\pm$ 0.050 & 0.9391 $\pm$ 0.053 \\
DeiT3-Medium       & 41.5M &  0.9575 $\pm$ 0.033 & 0.960 $\pm$ 0.031 & 0.958 $\pm$ 0.033 & 0.956 $\pm$ 0.035 \\
ViT-Base                & 88.7M &  0.9575 $\pm$ 0.035 & 0.9578 $\pm$ 0.033 & 0.9562 $\pm$ 0.036 & 0.9542 $\pm$ 0.036 \\
EfficientViT-L1          & 58.9M &  \underline{0.9725 $\pm$ 0.018} & \underline{0.9727 $\pm$ 0.017} & \underline{0.9744 $\pm$ 0.020} & \underline{0.9714 $\pm$ 0.019} \\
CoAtNet-1              & 41.7M & 0.9675 $\pm$ 0.037 & 0.971 $\pm$ 0.035 & 0.9671 $\pm$ 0.037 & 0.967 $\pm$ 0.037 \\
\textbf{EfficientViT-CAGA}            & \underline{36.8M} & \textbf{0.98 $\pm$ 0.0229} & \textbf{0.9789 $\pm$ 0.025} & \textbf{0.9795 $\pm$ 0.026} & \textbf{0.9781 $\pm$ 0.025} \\
\bottomrule
\end{tabular}

\vspace{5mm}

\centering
\setlength{\tabcolsep}{9pt}
\renewcommand{\arraystretch}{1.3}

\begin{tabular}{l|cccc}
\toprule
\multicolumn{1}{l}{Architectures} &
\multicolumn{4}{c}{Classwise Accuracy} \\
\cmidrule(lr){2-5}
\multicolumn{1}{c}{} & Mpox & Normal & Chickenpox & Acne \\
\midrule
ResNet-101                 & 0.8667 $\pm$ 0.116     & 0.8326 $\pm$ 0.131        & 0.8639 $\pm$ 0.126             & \textbf{1.0 $\pm$ 0.0}\\
MobileNetV3-Large          & 0.9063 $\pm$ 0.122     & 0.9205 $\pm$ 0.130        & \underline{0.9468 $\pm$ 0.093}             & \textbf{1.0 $\pm$ 0.0}\\
DeiT3-Medium               & 0.9538 $\pm$ 0.065     & 0.9707 $\pm$ 0.064        & 0.9188 $\pm$ 0.120              & \textbf{1.0 $\pm$ 0.0}\\
ViT-Base                        & 0.9180 $\pm$ 0.081     & \textbf{0.9923 $\pm$ 0.024}       & 0.9207 $\pm$ 0.118             & \textbf{1.0 $\pm$ 0.0}\\
EfficientViT-L1               & 0.9517 $\pm$ 0.066     & 0.9718 $\pm$ 0.062        & \textbf{0.9675 $\pm$ 0.0707}              & \textbf{1.0 $\pm$ 0.0}\\
CoAtNet-1                    & \textbf{0.9700 $\pm$ 0.051}     & \underline{0.9818 $\pm$ 0.057}        & 0.9325 $\pm$ 0.101             & \textbf{1.0 $\pm$ 0.0}\\
\textbf{EfficientViT-CAGA}    & \underline{0.9690 $\pm$ 0.065}     & 0.9789 $\pm$ 0.045        & \textbf{0.9675 $\pm$ 0.071}              & \textbf{1.0 $\pm$ 0.0}\\
\bottomrule
\end{tabular}
\label{Table : Mcsi}
\end{table*}

%this architectural approach offers three key advantages. First, it mitigates redundancy by augmenting inputs with previously processed head outputs, so each subsequent computation builds upon earlier representations rather than processing them independently. Second, it achieves parameter efficiency through head splitting, reducing the total parameter count by a factor of $n$. Third, it effectively increases model depth through sequential processing.

The processed attention heads are concatenated and projected back to the original input dimensions. A residual connection is then applied, linking the transformed representation with the initial input.
\begin{equation}
\hat{\textit{X}} = \textit{X} + \text{Proj}(\textit{Concat}[\hat{\textit{X}_i}]_{i=1:n})
\end{equation}
This final step ensures that the cascaded attention representation maintains a direct connection to the original input, preserving initial features while incorporating the hierarchically refined representations from the CGA and CAA mechanisms.

\begin{table}
\centering

\caption{FLOPs of models compared in this study}
\setlength{\tabcolsep}{7pt}
\renewcommand{\arraystretch}{1.2}
\begin{tabular}{l|cc}
\toprule
\textbf{Architecture}    & \textbf{FLOPs} \\
\midrule
ResNet-101 \cite{he2015deepresiduallearningimage}             & 7.86G \\
MobileNetV3-Large\cite{howard2019searchingmobilenetv3}         & \textbf{0.22G} \\
DeiT3-Medium\cite{touvron2022deitiiirevengevit}             & 7.52G \\
ViT-Base\cite{dosovitskiy2021imageworth16x16words}                     & 16.86G \\
EfficientViT-L1\cite{cai2024efficientvitmultiscalelinearattention}      & 5.27G \\
CoAtNet-1 \cite{dai2021coatnetmarryingconvolutionattention}              & 7.59G \\
\textbf{EfficientViT-CAGA}     & \underline{4.86G} \\
\bottomrule
\end{tabular}
\label{table : models and Flops}
\vspace{-10pt}
\end{table}

\begin{table*} [ht]
\centering

\caption{Comparison with SOTA Image Classification Models on MSLD Dataset} 

\setlength{\tabcolsep}{7pt}
\renewcommand{\arraystretch}{1.3}
\begin{tabular}{l|ccccccc}
\toprule
\multicolumn{1}{l}{Architectures} &
\multicolumn{1}{c}{Parameters} &
\multicolumn{1}{c}{Accuracy} &
\multicolumn{1}{c}{Precision} &
\multicolumn{1}{c}{Recall} &
\multicolumn{1}{c}{F1 Score} &
\multicolumn{2}{c}{Classwise Accuracy} \\
\cmidrule(lr){7-8}
\multicolumn{6}{c}{} & Mpox & Others \\

\midrule
ResNet-101           & 47.1M & 0.9593 & 0.9592 & 0.9586 & 0.9589 & 0.956 & 0.960 \\
MobileNetV3-Large     & \textbf{6.8M} &  0.9781 & 0.9782 & 0.9776 & 0.9779 & 0.979 & \underline{0.9974}\\
DeiT3-Medium         & 38.8M & \underline{0.9938} & \underline{0.9937} & 0.9937 & 0.9937 & \underline{0.993} & 0.994 \\
ViT-Base                  & 86.5M &  \underline{0.9938} & 0.993 & \underline{0.994} & \underline{0.994} & 0.986 & \textbf{1.0}\\
EfficientViT-L1          & 42M &  0.9875 & 0.987 & 0.988 & 0.987& 0.979& 0.994\\
CoAtNet-1              & 41.7M &  0.9906 & 0.99 & 0.99 & 0.99& 0.986 & 0.994 \\
\textbf{EfficientViT-CAGA}            & \underline{36.8M} &  \textbf{0.9969} & \textbf{0.997} & \textbf{0.997} & \textbf{0.997}& \textbf{1.0} & 0.994\\
\bottomrule
\end{tabular}
\label{Table : MSLD}

\end{table*}

\section{Experimental Results and Analysis}
This section introduces the dataset and implementation details, followed by a comparison of the proposed approach with state-of-the-art architectures. The proposed methodology is compared to six models, which are shown in Table \ref{table : models and Flops}. All models mentioned in Table \ref{table : models and Flops} are pre-trained on ImageNet, except for the classifier head which was randomly initialized following common practice. In contrast, for EfficientViT-CAGA, only the EfficientViT-L1 convolutional backbone is pre-trained, while the DSConv and CAGA module are initialized using the Xavier uniform method.

\subsection{Datasets}
To evaluate the performance of the proposed method, I selected three benchmark datasets. All images in the datasets are resized to $224 \times 224$.

The \textbf{Mpox Close Skin Image Dataset (MCSI)}\cite{CAMPANA2024101874} comprises four classes: Mpox, Chickenpox, Acne, and Normal, with each class containing 100 images.

The \textbf{Monkeypox Skin Images Dataset (MSID)}\cite{bala2023monkeynet} contains 770 total images distributed across four classes: Mpox, Chickenpox, Measles, and Acne. 

The \textbf{Monkeypox Skin Lesion Dataset (MSLD)}\cite{Nafisa2022}\cite{Nafisa2023} is a binary classification dataset with two classes: Mpox and Others (comprising Chickenpox and Measles). The total dataset size is 228 image.

\subsection{Implementation Details}

I utilize the ImageNet pre-trained EfficientViT-L1 \cite{cai2024efficientvitmultiscalelinearattention} as the backbone architecture, with the head initialized randomly. The implementation is done using PyTorch\cite{paszke2019pytorchimperativestylehighperformance} and trained on Tesla T4 GPUs. For pre-trained models I use Pytorch Image Models\cite{rw2019timm}. I employ the AdamW optimizer with a learning rate on the order of $10^{-5}$. The learning rate was adjusted during training using an exponential decay scheduler (ExponentialLR) with a multiplicative factor (gamma) of 0.95 per epoch. Training is regulated by an Early Stopping mechanism to prevent overfitting.

I use Z-score normalization to standardize the dataset. This is applied to all three datasets, with the mean and standard deviation computed using only the training data to avoid data leakage, and then applied consistently across the training, validation, and test sets.

For the CAA module, I utilize dilation rates $d \in \{1, 2, 3\}$, which are adjusted for each dataset, across a total of 3 attention heads. Each attention head has an embedding dimension of 16, while the dimensions for the query, key, and value embeddings ($d_{qkv}$) are set to 8. The weights are initialized using the Xavier uniform method with a fixed seed of 82.

% %\begin{table*} [h]
% %\centering
% \caption{Performance Metrics on MSID Dataset}
% \label{tab:metrics}
% \setlength{\tabcolsep}{7pt}
% \renewcommand{\arraystretch}{1.3}
% \begin{tabular}{l|ccccc}
% \toprule
% \multicolumn{1}{l}{Architectures} &
% \multicolumn{1}{|c}{Parameters} &
% \multicolumn{1}{c}{Accuracy} &
% \multicolumn{1}{c}{Precision} &
% \multicolumn{1}{c}{Recall} &
% \multicolumn{1}{c}{F1 Score} \\
% \midrule
% ResNet-101           & 43.5M & 0.896     & 0.8633        & 0.8782    & 0.8685 \\
% MobileNetV3-Large     & 5.5M & 0.9481 & 0.94  & 0.927 & 0.9327 \\
% DeiT3-Medium         & 38.8M & 0.9481 & 0.9159  & 0.9308 & 0.9229 \\
% ViT-Base                  & 86.5M & 0.9286 & 0.9097  & 0.9154 & 0.9113  \\
% EfficientViT          & 58.9M & 0.9481 & 0.9082 & 0.9273 & 0.9165 \\
% CoAtNet-1              & 41.7M & 0.9545 & 0.9201  & 0.9454 & 0.9317 \\
% EfficientViT + CAGA            & 37.8M & 0.9545 & 0.9523 & 0.9376  & 0.9446 \\
% \bottomrule
% \end{tabular}
% \end{table*}

\begin{figure*}[ht]
    \centering
    
    % First row - Class A
    \begin{minipage}{\textwidth}
        \centering
        % Add vspace to move the label down to center it with the images
        \raisebox{0cm}{\rotatebox[origin=c]{90}{Mpox}} \hspace{0.1cm}
        \begin{minipage}{0.14\textwidth}
            \includegraphics[width=\textwidth]{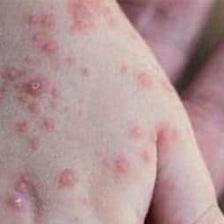}
        \end{minipage}%
        \hspace{0.01cm}
        \begin{minipage}{0.14\textwidth}
            \includegraphics[width=\textwidth]{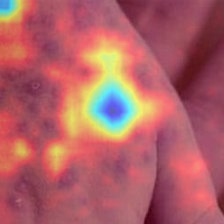}
        \end{minipage}%
        \hspace{0.2cm}
        \begin{minipage}{0.14\textwidth}
            \includegraphics[width=\textwidth]{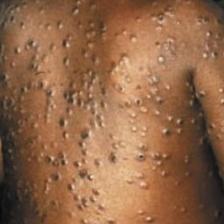}
        \end{minipage}%
        \hspace{0.01cm}
        \begin{minipage}{0.14\textwidth}
            \includegraphics[width=\textwidth]{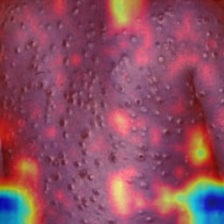}
        \end{minipage}%
        \hspace{0.2cm}
        \begin{minipage}{0.14\textwidth}
            \includegraphics[width=\textwidth]{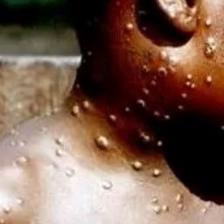}
        \end{minipage}%
        \hspace{0.01cm}
        \begin{minipage}{0.14\textwidth}
            \includegraphics[width=\textwidth]{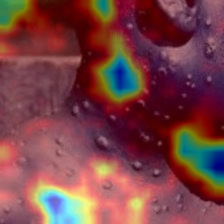}
        \end{minipage}
    \end{minipage}
    
    \vspace{0.1cm}
    
    % Second row - Class B
    \begin{minipage}{\textwidth}
        \centering
        % Add vspace to move the label down to center it with the images
        \raisebox{0cm}{\rotatebox[origin=c]{90}{Others}} \hspace{0.1cm}
        \begin{minipage}{0.14\textwidth}
            \includegraphics[width=\textwidth]{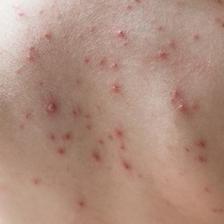}
        \end{minipage}%
        \hspace{0.01cm}
        \begin{minipage}{0.14\textwidth}
            \includegraphics[width=\textwidth]{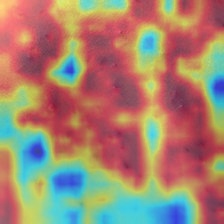}
        \end{minipage}%
        \hspace{0.2cm}
        \begin{minipage}{0.14\textwidth}
            \includegraphics[width=\textwidth]{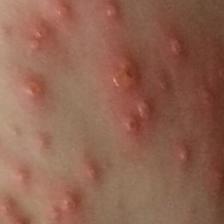}
        \end{minipage}%
        \hspace{0.01cm}
        \begin{minipage}{0.14\textwidth}
            \includegraphics[width=\textwidth]{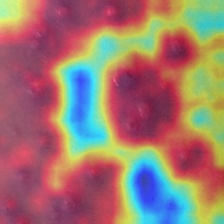}
        \end{minipage}%
        \hspace{0.2cm}
        \begin{minipage}{0.14\textwidth}
            \includegraphics[width=\textwidth]{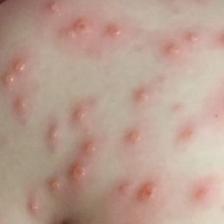}
        \end{minipage}%
        \hspace{0.01cm}
        \begin{minipage}{0.14\textwidth}
            \includegraphics[width=\textwidth]{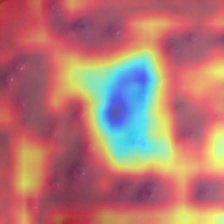}
        \end{minipage}
    \end{minipage}
    
    \caption{Original images and their corresponding Grad-CAM visualizations for Mpox and Others from the MSLD.}
    \label{fig:gradcam_comparison}
\vspace{-10pt}
\end{figure*}

\subsection{Results}
The bolded results for each metric indicate the highest score among all models, while the second-highest scores are underlined. For parameters and FLOPs, the smallest value is bolded, and the second smallest is underlined.
\subsubsection{MCSI}

The experimental evaluation was conducted using 10-fold cross-validation, following the methodology proposed by \cite{CAMPANA2024101874}. For each fold, the dataset was partitioned into three distinct sets: a training set comprising 288 images, a validation set of 72 images, and a test set containing 40 images. Although runtime data augmentation techniques were applied to the training set, these modifications led to a marginal decrease in performance metrics across all evaluated models, which is consistent with the findings in \cite{CAMPANA2024101874}. For all models, the loss function used was focal loss with class-specific weights (alpha). While Focal Loss\cite{lin2018focallossdenseobject} is traditionally employed to address class imbalance, experiments showed it significantly improved model performance despite having evenly distributed classes. 

Table \ref{Table : Mcsi} reports the comparison of EfficientViT-CAGA with other state-of-the-art models. EfficientViT-CAGA achieves remarkable accuracy while being computationally efficient, with only 4.86G FLOPs, which is 3.5$\times$ less than ViT-Base. Compared to the backbone EfficientViT-L1, proposed EfficientViT-CAGA achieves a significant 37.5\% reduction in total parameters while improving performance across all evaluation metrics. 

\subsubsection{MSLD}
As MSLD contains fewer than 250 total images, \cite{Nafisa2022} employed data augmentation techniques to expand the dataset. These included standard methods such as rotation, translation, reflection, shear, hue adjustment, saturation, contrast and brightness jitter, noise addition, and scaling.

These augmentations expanded the dataset to over 1,428 images for the Mpox class and 1,764 images for the Others class. The augmented dataset was then divided into training, validation, and testing sets using a 70:20:10 split ratio with standard ImageNet normalization applied. 

Table \ref{Table : MSLD} presents a comparison between EfficientViT-CAGA and SOTA models for image classification on the MSLD dataset. Similar to the results on MCSI, EfficientViT-CAGA outperforms competing models across almost all evaluation metrics. 
It achieves this with approximately 12\% fewer parameters compared to both EfficientViT-L1 and CoAtNet-1, both of which show competitive results. Notably, EfficientViT-CAGA is the only model to achieve 100\% accuracy on the Mpox class.

\begin{table}
\centering

\caption{Ablation Study Results}
\setlength{\tabcolsep}{7pt}
\renewcommand{\arraystretch}{1.2}
\begin{tabular}{c|c|c|c|c}
\toprule
\multicolumn{1}{l}{Cascading in CAA } &
\multicolumn{1}{|l}{CAA} &
\multicolumn{1}{|l}{CGA} &
\multicolumn{1}{|c}{Params} &
\multicolumn{1}{|c}{Accuracy} \\
\midrule
 \checkmark & \checkmark &  & 36.7M &0.77\\
 & \checkmark &\checkmark& 36.8M & 0.815\\
\checkmark & \checkmark &\checkmark& 36.8M & 0.8175\\
 
\bottomrule
\end{tabular}
\label{ablation}
\vspace{-10pt}
\end{table}

\subsubsection{MSID}

A total of 616 images were used for training and 154 for testing, following the standard 80-20 split. Similar to MCSI, I applied augmentation to the MSID training data, but it led to worse performance.

Table \ref{table:msid} presents a consolidated comparison of EfficientViT-CAGA with SOTA models on MSID. The overall results show marginal variations across different architectures, though a significant distinction is observed in class-wise accuracy. EfficientViT-CAGA achieves balanced performance across all classes, with a standard deviation of only 0.8\% in class-wise accuracy. In comparison, other architectures tend to overfit certain classes. For instance, CoAtNet-1 achieves high accuracy of 0.9821 for Mpox and 1.0 for Normal classes but drops significantly to 0.81 for Chickenpox and 0.89 for Measles, resulting in a higher standard deviation of 7.7\%. Compared to the backbone, the proposed model achieves a 35.8\% reduction in parameters while outperforming it in accuracy. Additionally, it has 2.2$\times$ fewer parameters than ViT-Base, yet demonstrates a significant improvement in accuracy.

\subsection{Ablation Study}\label{SCM}
I conducted a rigorous ablation study to evaluate the Cascaded Atrous Attention and Cascaded Group Attention modules on the MCSI dataset. The models tested were initialized with random weights to eliminate pre-training biases. The experiments are conducted using three attention heads with fixed dilation rates ($d \in \{1,2\}$) and a fixed classifier head to ensure consistent experimental conditions.

In this ablation study, I first evaluate the impact of the cascading mechanism within the CAA module by eliminating cascading between dilations. Second, I quantify the performance contribution of the CGA module by excluding it, retaining only the CAA module.

Table \ref{ablation} summarizes the results of the ablation study. The accuracy is reported as the average across 10-fold cross-validation. The findings indicate that there are no notable differences in the number of parameters, and the use of CGA and the cascading component contributes to a slight improvement in the model's performance.

Alongside this, I compute Grad-CAM heatmaps to provide a more interpretable approach to understanding what the model focuses on when making a decision. By generating these heatmaps, I visualize the key regions of the input images that contribute to the model's classification. This approach is applied to MSLD, and the corresponding visualizations are shown in Fig. \ref{fig:gradcam_comparison}.

\def\checkmark{\tikz\fill[scale=0.4](0,.35) -- (.25,0) -- (1,.7) -- (.25,.15) -- cycle;}

\section{Conclusion}

In this work, I proposed the Cascaded Atrous Group Attention architecture, which combines the Cascaded Atrous Attention and the Cascaded Group Attention mechanism to improve the efficiency and accuracy for Mpox classification. CAA achieved multi-scale representation using dilated convolutions and cascaded these outputs to enhance spatial and contextual information. CAA was encapsulated in a CGA, originally designed to address the redundancy in Multi-Head Self-Attention. Experimental results show that the approach highlighted state-of-the-art results on three datasets. In future work, I plan to extend the application of CAGA to broader tasks, such as disease segmentation and object detection, where multi-scale feature representation and efficient attention mechanisms could drive substantial improvements. Integrating self-supervised learning techniques with CAGA could improve its performance, especially when applied to limited or imbalanced medical datasets.

\bibliographystyle{unsrt}
\bibliography{ref.bib}

\begin{thebibliography}{10}

\bibitem{WHO2022}
World~Health Organization.
\newblock Multi-country monkeypox outbreak in non-endemic countries, May 2022.

\bibitem{:/content/10.2807/1560-7917.ES.2023.28.12.2200772}
Catharina~E van Ewijk, Fuminari Miura, Gini van Rijckevorsel, Henry~JC de~Vries, Matthijs~RA Welkers, Oda~E van~den Berg, Ingrid~HM Friesema, Patrick~R van~den Berg, Thomas Dalhuisen, Jacco Wallinga, Diederik Brandwagt, Brigitte~AGL van Cleef, Harry Vennema, Bettie Voordouw, Marion Koopmans, Annemiek~A van~der Eijk, Corien~M Swaan, Margreet~JM te~Wierik, Tjalling Leenstra, Eline Op~de Coul, Eelco Franz, and the Dutch Mpox Response~Team.
\newblock Mpox outbreak in the netherlands, 2022: public health response, characteristics of the first 1,000 cases and protection of the first-generation smallpox vaccine.
\newblock {\em Eurosurveillance}, 28(12), 2023.

\bibitem{v15061384}
Kassiani Mellou, Kyriaki Tryfinopoulou, Styliani Pappa, Kassiani Gkolfinopoulou, Sofia Papanikou, Georgia Papadopoulou, Evangelia Vassou, Evangelia-Georgia Kostaki, Kalliopi Papadima, Elissavet Mouratidou, Maria Tsintziloni, Nikolaos Siafakas, Zoi Florou, Antigoni Katsoulidou, Spyros Sapounas, George Sourvinos, Spyridon Pournaras, Efthymia Petinaki, Maria Goula, Vassilios Paparizos, Anna Papa, Theoklis Zaoutis, and Dimitrios Paraskevis.
\newblock Overview of mpox outbreak in greece in 2022–2023: Is it over?
\newblock {\em Viruses}, 15(6), 2023.

\bibitem{borges2023viral}
Vítor Borges, Maria~P. Duque, Joana~V. Martins, Paula Vasconcelos, Rita Ferreira, Duarte Sobral, Ana Pelerito, Isabel~L. de~Carvalho, Marta~S. Núncio, Maria~J. Borrego, Cornelia Roemer, Richard~A. Neher, Megan O'Driscoll, Raquel Rocha, Sofia Lopo, Ricardo Neves, Patricia Palminha, Liliana Coelho, Andreia Nunes, Joana Isidro, and João~Paulo Gomes.
\newblock Viral genetic clustering and transmission dynamics of the 2022 mpox outbreak in portugal.
\newblock {\em Nature Medicine}, 29(10):2509--2517, 2023.

\bibitem{:/content/10.2807/1560-7917.ES.2023.28.50.2200923}
Catarina Krug, Arnaud Tarantola, Emilie Chazelle, Erica Fougère, Annie Velter, Anne Guinard, Yvan Souares, Anna Mercier, Céline François, Katia Hamdad, Laetitia Tan-Lhernould, Anita Balestier, Hana Lahbib, Nicolas Etien, Pascale Bernillon, Virginie De~Lauzun, Julien Durand, Myriam Fayad, Investigation Team, Henriette De~Valk, François Beck, Didier Che, Bruno Coignard, Florence Lot, and Alexandra Mailles.
\newblock Mpox outbreak in france: epidemiological characteristics and sexual behaviour of cases aged 15 years or older, 2022.
\newblock {\em Eurosurveillance}, 28(50), 2023.

\bibitem{owid-mpox}
Edouard Mathieu, Fiona Spooner, Saloni Dattani, Hannah Ritchie, and Max Roser.
\newblock Mpox.
\newblock {\em Our World in Data}, 2022.
\newblock https://ourworldindata.org/mpox.

\bibitem{silva2023clinical}
Shania J. R.~D. Silva, Alain Kohl, Luis Pena, and Keith Pardee.
\newblock Clinical and laboratory diagnosis of monkeypox (mpox): Current status and future directions.
\newblock {\em iScience}, 26(6):106759, 2023.

\bibitem{vaccines11061093}
Shriyansh Srivastava, Sachin Kumar, Shagun Jain, Aroop Mohanty, Neeraj Thapa, Prabhat Poudel, Krishna Bhusal, Zahraa~Haleem Al-qaim, Joshuan~J. Barboza, Bijaya~Kumar Padhi, and Ranjit Sah.
\newblock The global monkeypox (mpox) outbreak: A comprehensive review.
\newblock {\em Vaccines}, 11(6), 2023.

\bibitem{prasad2023dermatologic}
Surbhi Prasad, Claudia Galvan~Casas, Andrew~G. Strahan, Lloyd~C. Fuller, Kathryn Peebles, Alessandra Carugno, Kenneth~S. Leslie, Jennifer~L. Harp, Tudor Pumnea, David~E. McMahon, Michael Rosenbach, Jonathan~E. Lubov, George Chen, Lynn~P. Fox, Allison McMillen, Hyun~W. Lim, Alexander~J. Stratigos, Thomas~A. Cronin, Marc~D. Kaufmann, Gerald~J. Hruza, and Eric~E. Freeman.
\newblock A dermatologic assessment of 101 mpox (monkeypox) cases from 13 countries during the 2022 outbreak: Skin lesion morphology, clinical course, and scarring.
\newblock {\em Journal of the American Academy of Dermatology}, 88(5):1066--1073, 2023.

\bibitem{esteva2017dermatologist}
Andre Esteva, Brett Kuprel, Roberto~A Novoa, Justin Ko, Susan~M Swetter, Helen~M Blau, and Sebastian Thrun.
\newblock Dermatologist-level classification of skin cancer with deep neural networks.
\newblock {\em nature}, 542(7639):115--118, 2017.

\bibitem{haenssle2018man}
Holger~A Haenssle, Christine Fink, Roland Schneiderbauer, Ferdinand Toberer, Timo Buhl, Andreas Blum, Aadi Kalloo, A~Ben~Hadj Hassen, Luc Thomas, Alexander Enk, et~al.
\newblock Man against machine: diagnostic performance of a deep learning convolutional neural network for dermoscopic melanoma recognition in comparison to 58 dermatologists.
\newblock {\em Annals of oncology}, 29(8):1836--1842, 2018.

\bibitem{thomsen2020systematic}
Kenneth Thomsen, Lars Iversen, Therese~Louise Titlestad, and Ole Winther.
\newblock Systematic review of machine learning for diagnosis and prognosis in dermatology.
\newblock {\em Journal of Dermatological Treatment}, 31(5):496--510, 2020.

\bibitem{7893267}
Adria Romero~Lopez, Xavier Giro-i Nieto, Jack Burdick, and Oge Marques.
\newblock Skin lesion classification from dermoscopic images using deep learning techniques.
\newblock In {\em 2017 13th IASTED International Conference on Biomedical Engineering (BioMed)}, pages 49--54, 2017.

\bibitem{chadaga2023application}
K.~Chadaga, S.~Prabhu, N.~Sampathila, S.~Nireshwalya, S.~S. Katta, R.~S. Tan, and U.~R. Acharya.
\newblock Application of artificial intelligence techniques for monkeypox: A systematic review.
\newblock {\em Diagnostics (Basel, Switzerland)}, 13(5):824, 2023.

\bibitem{asif2024ai}
S.~Asif, M.~Zhao, Y.~Li, et~al.
\newblock Ai-based approaches for the diagnosis of mpox: Challenges and future prospects.
\newblock {\em Arch Computational Methods in Engineering}, 31:3585--3617, August 2024.
\newblock Received: 28 October 2023; Accepted: 04 February 2024; Published: 26 March 2024.

\bibitem{jaradat2023automated}
A.S. Jaradat, R.E. Al~Mamlook, N.~Almakayeel, N.~Alharbe, A.S. Almuflih, A.~Nasayreh, H.~Gharaibeh, M.~Gharaibeh, A.~Gharaibeh, and H.~Bzizi.
\newblock Automated monkeypox skin lesion detection using deep learning and transfer learning techniques.
\newblock {\em International Journal of Environmental Research and Public Health}, 20(5):4422, 2023.

\bibitem{CAMPANA2024101874}
Mattia~Giovanni Campana, Marco Colussi, Franca Delmastro, Sergio Mascetti, and Elena Pagani.
\newblock A transfer learning and explainable solution to detect mpox from smartphones images.
\newblock {\em Pervasive and Mobile Computing}, 98:101874, 2024.

\bibitem{10491259}
Avi~Deb Raha, Mrityunjoy Gain, Rameswar Debnath, Apurba Adhikary, Yu~Qiao, Md.~Mehedi Hassan, Anupam~Kumar Bairagi, and Sheikh Mohammed~Shariful Islam.
\newblock Attention to monkeypox: An interpretable monkeypox detection technique using attention mechanism.
\newblock {\em IEEE Access}, 12:51942--51965, 2024.

\bibitem{thieme2023deep}
A.H. Thieme, Y.~Zheng, G.~Machiraju, et~al.
\newblock A deep-learning algorithm to classify skin lesions from mpox virus infection.
\newblock {\em Nature Medicine}, 29:738--747, March 2023.
\newblock Received: 05 August 2022; Accepted: 19 January 2023; Published: 02 March 2023.

\bibitem{cai2024efficientvitmultiscalelinearattention}
Han Cai, Junyan Li, Muyan Hu, Chuang Gan, and Song Han.
\newblock Efficientvit: Multi-scale linear attention for high-resolution dense prediction, 2024.

\bibitem{bala2023monkeynet}
Diponkor Bala, Md~Shamim Hossain, Mohammad~Alamgir Hossain, Md~Ibrahim Abdullah, Md~Mizanur Rahman, Balachandran Manavalan, Naijie Gu, Mohammad~S Islam, and Zhangjin Huang.
\newblock Monkeynet: A robust deep convolutional neural network for monkeypox disease detection and classification.
\newblock {\em Neural Networks}, 161:757--775, 2023.

\bibitem{Nafisa2022}
Shams~Nafisa Ali, Md.~Tazuddin Ahmed, Joydip Paul, Tasnim Jahan, S.~M.~Sakeef Sani, Nawshaba Noor, and Taufiq Hasan.
\newblock Monkeypox skin lesion detection using deep learning models: A preliminary feasibility study.
\newblock {\em arXiv preprint arXiv:2207.03342}, 2022.

\bibitem{Nafisa2023}
Shams~Nafisa Ali, Md.~Tazuddin Ahmed, Tasnim Jahan, Joydip Paul, S.~M.~Sakeef Sani, Nawshaba Noor, Anzirun~Nahar Asma, and Taufiq Hasan.
\newblock A web-based mpox skin lesion detection system using state-of-the-art deep learning models considering racial diversity.
\newblock {\em arXiv preprint arXiv:2306.14169}, 2023.

\bibitem{szegedy2014goingdeeperconvolutions}
Christian Szegedy, Wei Liu, Yangqing Jia, Pierre Sermanet, Scott Reed, Dragomir Anguelov, Dumitru Erhan, Vincent Vanhoucke, and Andrew Rabinovich.
\newblock Going deeper with convolutions, 2014.

\bibitem{NIPS2012_c399862d}
Alex Krizhevsky, Ilya Sutskever, and Geoffrey~E Hinton.
\newblock Imagenet classification with deep convolutional neural networks.
\newblock In F.~Pereira, C.J. Burges, L.~Bottou, and K.Q. Weinberger, editors, {\em Advances in Neural Information Processing Systems}, volume~25. Curran Associates, Inc., 2012.

\bibitem{xie2017aggregatedresidualtransformationsdeep}
Saining Xie, Ross Girshick, Piotr Dollár, Zhuowen Tu, and Kaiming He.
\newblock Aggregated residual transformations for deep neural networks, 2017.

\bibitem{he2015deepresiduallearningimage}
Kaiming He, Xiangyu Zhang, Shaoqing Ren, and Jian Sun.
\newblock Deep residual learning for image recognition, 2015.

\bibitem{chollet2017xceptiondeeplearningdepthwise}
François Chollet.
\newblock Xception: Deep learning with depthwise separable convolutions, 2017.

\bibitem{tan2020efficientnetrethinkingmodelscaling}
Mingxing Tan and Quoc~V. Le.
\newblock Efficientnet: Rethinking model scaling for convolutional neural networks, 2020.

\bibitem{vaswani2023attentionneed}
Ashish Vaswani, Noam Shazeer, Niki Parmar, Jakob Uszkoreit, Llion Jones, Aidan~N. Gomez, Lukasz Kaiser, and Illia Polosukhin.
\newblock Attention is all you need, 2023.

\bibitem{dosovitskiy2021imageworth16x16words}
Alexey Dosovitskiy, Lucas Beyer, Alexander Kolesnikov, Dirk Weissenborn, Xiaohua Zhai, Thomas Unterthiner, Mostafa Dehghani, Matthias Minderer, Georg Heigold, Sylvain Gelly, Jakob Uszkoreit, and Neil Houlsby.
\newblock An image is worth 16x16 words: Transformers for image recognition at scale, 2021.

\bibitem{raghu2022visiontransformerslikeconvolutional}
Maithra Raghu, Thomas Unterthiner, Simon Kornblith, Chiyuan Zhang, and Alexey Dosovitskiy.
\newblock Do vision transformers see like convolutional neural networks?, 2022.

\bibitem{park2022visiontransformerswork}
Namuk Park and Songkuk Kim.
\newblock How do vision transformers work?, 2022.

\bibitem{chen2022visiontransformersoutperformresnets}
Xiangning Chen, Cho-Jui Hsieh, and Boqing Gong.
\newblock When vision transformers outperform resnets without pre-training or strong data augmentations, 2022.

\bibitem{guo2022cmtconvolutionalneuralnetworks}
Jianyuan Guo, Kai Han, Han Wu, Yehui Tang, Xinghao Chen, Yunhe Wang, and Chang Xu.
\newblock Cmt: Convolutional neural networks meet vision transformers, 2022.

\bibitem{liu2022convnet2020s}
Zhuang Liu, Hanzi Mao, Chao-Yuan Wu, Christoph Feichtenhofer, Trevor Darrell, and Saining Xie.
\newblock A convnet for the 2020s, 2022.

\bibitem{graham2021levitvisiontransformerconvnets}
Ben Graham, Alaaeldin El-Nouby, Hugo Touvron, Pierre Stock, Armand Joulin, Hervé Jégou, and Matthijs Douze.
\newblock Levit: a vision transformer in convnet's clothing for faster inference, 2021.

\bibitem{liu2021swintransformerhierarchicalvision}
Ze~Liu, Yutong Lin, Yue Cao, Han Hu, Yixuan Wei, Zheng Zhang, Stephen Lin, and Baining Guo.
\newblock Swin transformer: Hierarchical vision transformer using shifted windows, 2021.

\bibitem{dai2021coatnetmarryingconvolutionattention}
Zihang Dai, Hanxiao Liu, Quoc~V. Le, and Mingxing Tan.
\newblock Coatnet: Marrying convolution and attention for all data sizes, 2021.

\bibitem{WONGVIBULSIN2023283}
Shannon Wongvibulsin and Adewole~S. Adamson.
\newblock Deep learning for mpox: Advances, challenges, and opportunities.
\newblock {\em Med}, 4(5):283--284, 2023.

\bibitem{ASIF2023342}
Sohaib Asif, Ming Zhao, Fengxiao Tang, Yusen Zhu, and Baokang Zhao.
\newblock Metaheuristics optimization-based ensemble of deep neural networks for mpox disease detection.
\newblock {\em Neural Networks}, 167:342--359, 2023.

\bibitem{sandler2019mobilenetv2invertedresidualslinear}
Mark Sandler, Andrew Howard, Menglong Zhu, Andrey Zhmoginov, and Liang-Chieh Chen.
\newblock Mobilenetv2: Inverted residuals and linear bottlenecks, 2019.

\bibitem{ioffe2015batchnormalizationacceleratingdeep}
Sergey Ioffe and Christian Szegedy.
\newblock Batch normalization: Accelerating deep network training by reducing internal covariate shift, 2015.

\bibitem{liu2023efficientvitmemoryefficientvision}
Xinyu Liu, Houwen Peng, Ningxin Zheng, Yuqing Yang, Han Hu, and Yixuan Yuan.
\newblock Efficientvit: Memory efficient vision transformer with cascaded group attention, 2023.

\bibitem{deeplab}
Liang-Chieh Chen, George Papandreou, Iasonas Kokkinos, Kevin Murphy, and Alan~L. Yuille.
\newblock Deeplab: Semantic image segmentation with deep convolutional nets, atrous convolution, and fully connected crfs.
\newblock {\em IEEE Transactions on Pattern Analysis and Machine Intelligence}, 40(4):834--848, 2018.

\bibitem{gao2019representationdegenerationproblemtraining}
Jun Gao, Di~He, Xu~Tan, Tao Qin, Liwei Wang, and Tie-Yan Liu.
\newblock Representation degeneration problem in training natural language generation models, 2019.

\bibitem{touvron2021goingdeeperimagetransformers}
Hugo Touvron, Matthieu Cord, Alexandre Sablayrolles, Gabriel Synnaeve, and Hervé Jégou.
\newblock Going deeper with image transformers, 2021.

\bibitem{zhou2021deepvitdeepervisiontransformer}
Daquan Zhou, Bingyi Kang, Xiaojie Jin, Linjie Yang, Xiaochen Lian, Zihang Jiang, Qibin Hou, and Jiashi Feng.
\newblock Deepvit: Towards deeper vision transformer, 2021.

\bibitem{zhou2021refinerrefiningselfattentionvision}
Daquan Zhou, Yujun Shi, Bingyi Kang, Weihao Yu, Zihang Jiang, Yuan Li, Xiaojie Jin, Qibin Hou, and Jiashi Feng.
\newblock Refiner: Refining self-attention for vision transformers, 2021.

\bibitem{chen2022principlediversitytrainingstronger}
Tianlong Chen, Zhenyu Zhang, Yu~Cheng, Ahmed Awadallah, and Zhangyang Wang.
\newblock The principle of diversity: Training stronger vision transformers calls for reducing all levels of redundancy, 2022.

\bibitem{howard2019searchingmobilenetv3}
Andrew Howard, Mark Sandler, Grace Chu, Liang-Chieh Chen, Bo~Chen, Mingxing Tan, Weijun Wang, Yukun Zhu, Ruoming Pang, Vijay Vasudevan, Quoc~V. Le, and Hartwig Adam.
\newblock Searching for mobilenetv3, 2019.

\bibitem{touvron2022deitiiirevengevit}
Hugo Touvron, Matthieu Cord, and Hervé Jégou.
\newblock Deit iii: Revenge of the vit, 2022.

\bibitem{paszke2019pytorchimperativestylehighperformance}
Adam Paszke, Sam Gross, Francisco Massa, Adam Lerer, James Bradbury, Gregory Chanan, Trevor Killeen, Zeming Lin, Natalia Gimelshein, Luca Antiga, Alban Desmaison, Andreas Köpf, Edward Yang, Zach DeVito, Martin Raison, Alykhan Tejani, Sasank Chilamkurthy, Benoit Steiner, Lu~Fang, Junjie Bai, and Soumith Chintala.
\newblock Pytorch: An imperative style, high-performance deep learning library, 2019.

\bibitem{rw2019timm}
Ross Wightman.
\newblock Pytorch image models.
\newblock \url{https://github.com/rwightman/pytorch-image-models}, 2019.

\bibitem{lin2018focallossdenseobject}
Tsung-Yi Lin, Priya Goyal, Ross Girshick, Kaiming He, and Piotr Dollár.
\newblock Focal loss for dense object detection, 2018.

\end{thebibliography}
\vspace{12pt}

\end{document}